\documentclass[aps,prl,10pt,twocolumn,superscriptaddress]{revtex4}
\usepackage{amsmath}
\usepackage{amssymb}
\usepackage{bbm}
\usepackage{graphicx}
\usepackage{framed}
\usepackage{color}
\usepackage{hyperref}
\usepackage{epstopdf}
\usepackage{dsfont}
\usepackage{amssymb}
\usepackage{amsmath}
\usepackage{amsthm}
\usepackage{wrapfig}
\usepackage{relsize}
\usepackage{bm}
\usepackage[latin1]{inputenc}
\usepackage{enumitem}
\usepackage{subfigure}

\newcommand{\red}[1]{{\color{red}}}

\usepackage{pifont}

\newcommand{\ba}{\begin{eqnarray}}
\newcommand{\ea}{\end{eqnarray}}

\newcommand{\identity}{\mathlarger{
\mathds{1}}}

\begin{document}

\title{Bridging thermodynamics and metrology in non-equilibrium Quantum Thermometry} 

\author{Vasco Cavina}
\affiliation{NEST, Scuola Normale Superiore and Istituto Nanoscienze-CNR, Piazza dei Cavalieri 7, I-56126, Pisa, Italy}

\author{Luca Mancino}
\affiliation{Dipartimento di Scienze, Universit\`{a} degli Studi Roma Tre, Via della Vasca Navale 84, 00146, Rome, Italy}

\author{Antonella De Pasquale}
\affiliation{Dipartimento di Fisica, Universit\`{a} di Firenze, Via G. Sansone 1, I-50019, Sesto Fiorentino (FI), Italy}
\affiliation{INFN Sezione di Firenze, via G.Sansone 1, I-50019 Sesto Fiorentino (FI), Italy}
\affiliation{NEST, Scuola Normale Superiore and Istituto Nanoscienze-CNR, Piazza dei Cavalieri 7, I-56126, Pisa, Italy}

\author{Ilaria Gianani}
\affiliation{Dipartimento di Scienze, Universit\`{a} degli Studi Roma Tre, Via della Vasca Navale 84, 00146, Rome, Italy}

\author{Marco Sbroscia}
\affiliation{Dipartimento di Scienze, Universit\`{a} degli Studi Roma Tre, Via della Vasca Navale 84, 00146, Rome, Italy}

\author{Robert I. Booth}
\affiliation{Dipartimento di Scienze, Universit\`{a} degli Studi Roma Tre, Via della Vasca Navale 84, 00146, Rome, Italy}
\affiliation{Institut de Physique, Sorbonne Universit\'{e}, 4 Place Jussieu, 75005, Paris, France}

\author{Emanuele Roccia}
\affiliation{Dipartimento di Scienze, Universit\`{a} degli Studi Roma Tre, Via della Vasca Navale 84, 00146, Rome, Italy}

\author{Roberto Raimondi}
\affiliation{Dipartimento di Matematica e Fisica, Universit\`{a} degli Studi Roma Tre, Via della Vasca Navale 84, 00146, Rome, Italy}

\author{Vittorio Giovannetti}
\affiliation{NEST, Scuola Normale Superiore and Istituto Nanoscienze-CNR, Piazza dei Cavalieri 7, I-56126, Pisa, Italy}

\author{Marco Barbieri}
\affiliation{Dipartimento di Scienze, Universit\`{a} degli Studi Roma Tre, Via della Vasca Navale 84, 00146, Rome, Italy}
\affiliation{Istituto Nazionale di Ottica - CNR, Largo Enrico Fermi 6, 50125, Florence, Italy}

\begin{abstract} 
Single-qubit thermometry presents the simplest tool to measure the temperature of thermal baths with reduced invasivity. At thermal equilibrium,  the temperature uncertainty is linked to the heat capacity of the qubit, however the best precision is achieved outside equilibrium condition. Here, we discuss a way to generalize this relation in a non-equilibrium regime, taking into account purely quantum effects such as coherence. We support our findings with an experimental photonic simulation. 
\end{abstract}

\maketitle

\paragraph{Introduction:--} 
Identifying strategies for improving 
the measurement precision by means of quantum resources is the purpose of Quantum Metrology \cite{Giovannetti06,Giovannetti11,Paris09}. In particular, through the Quantum Cram\'{e}r-Rao Bound (QCRB), it sets ultimate limits on the best accuracy attainable in the  estimation of unknown parameters  even when the latter are not associated with observable quantities.
These considerations have attracted an increasing attention in the field of quantum thermodynamics, where an accurate control of the temperature is highly demanding \cite{Hilt09, Williams11, Kliesch14, Vinjanampathy16,Millen16}. Besides the emergence of primary and secondary thermometers based on precisely machined microwave resonators \cite{Mohr05, Weng14}, recent efforts have been made aiming at measuring temperature at even smaller scales, where nanosize thermal baths are 
higly sensitives to disturbances induced by the probe
\cite{Stace10,Mann14,Mehboudi15,Depasquale16,
Depasquale17,DePalma17,Campbell117}.
Some paradigmatic examples of nanoscale thermometry involve nanomechanical resonators \cite{Brunelli11}, quantum harmonic oscillators \cite{Brunelli12} or atomic condensates \cite{Sabin14,Johnson16,Hohmann16} (also in conjunction with estimation of chemical potential \cite{Marzolino13}). In this context the analysis of quantum properties needs to be taken into account in order to establish, and eventually enhance, metrological precision \cite{Salvatori14,Jevtic15,Tham16,Mancino17,Correa17,Campbell18}. 

In a conventional  approach to thermometry,
  an external bath  ${\cal B}$  at thermal equilibrium
 is typically indirectly probed 
via an ancillary system, the thermometer  ${\cal S}$,  that 
is placed into weak-interaction with the former. 
 Assuming hence that the thermometer reaches the thermal equilibrium configuration without perturbing ${\cal B}$ too much, 
 the Einstein Theory of Fluctuations (ETF) can be used to characterize the sensitivity of the procedure  
in terms of the  heat capacity of ${\cal S}$ which represents its thermal susceptibility 
  to the perturbation imposed by the bath~\cite{Landau80,Falcioni11,Dicastro15}. 
  Since this last is an equilibrium property, one should not expect it to hold in non-equilibrium regimes. However thermometry schemes that do not need a full thermalization of the probe have been recognized to offer higher sensitivities in temperature estimation \cite{Correa15}. Thus, if on the one hand the QCRB can still be used as the proper tool to 
  gauge the measurement uncertainty on the bath temperature,
on the other hand establishing a direct link between this approach and the thermodynamic properties of the probe is still an open question.  Furthermore, the advantages pointed out in~\cite{Correa15}
 are conditional on precisely addressing the probe during its evolution, a task which might be demanding in real experiments \cite{Mancino17}.
 Here ${\cal S}$ is assumed to be a quantum system characterized by a local Hamiltonian $H$
that, after being initialized into some proper input state $\rho(0)$,  weakly interacts for some time $\tau$ with the bath ${\cal B}$ of assigned, but unknown, temperature $T$, before been measured. 
In this setting, we compare the performances of optimal estimation procedures with standard thermometry approaches: the temperature parameter $T$ is recovered by only monitoring the energy variation on ${\cal S}$ by its interaction with the bath. Then we derive a universal inequality that links metrological and thermodynamic quantities, ultimately discussing the optimal condition for its saturation.

In particular for the case where ${\cal S}$ is a two-level (qubit) system we show that optimality can be achieved for
a broad class of configurations which also include out of equilibrium scenarios for which ETF does not holds.
These results are also confirmed by an experiment where the proposed scheme is simulated via quantum photonics.

\paragraph{QCRB vs ETF:--} 
A direct application of the QCRB
~\cite{Paris09,Giovannetti11} to our setting 
 establishes that the  Mean Square Error (MSE)  $\Delta^2 T$ 
of any temperature estimation procedure, based on an arbitrary local measurement on ${\cal S}$,  is limited by the  inequality 
 $\Delta^2 T \geq 1/[M Q_{T}(\tau)]$. In this expression $M$ is the number of measurements one performs on the probe, while 
 $Q_{T}(\tau)$ is the  Quantum Fisher Information (QFI): a complex functional which only depends on the reduced density matrix $\rho(\tau)$  describing the state of  ${\cal S}$ after its interaction with ${\cal B}$ (see  below for details). 
Consider then the case where, as in the conventional thermometry approach,  the bath temperature is recovered by just measuring the mean energy $E_T(\tau)=\mbox{tr}[ H\rho(\tau)]$  of ${\cal S}$ and 
 inverting its functional dependence upon $T$. A simple application
 of  the error propagation formula reveals that in this scenario the associated MSE can be expressed as $\Delta^2 T = \Delta^2 E_T(\tau)/[{M C_T^2(\tau)}]$,
 where  $\Delta^2E (\tau)=\mbox{tr}[ (H-E_T(\tau))^2 \rho(\tau)]$ is the
 variance of $H$ on $\rho(\tau)$ we use to estimate  the uncertainty of the mean energy $E_T(\tau)$, and $C_T(\tau) = \partial_T E_T(\tau)$ is the partial derivative  of $E_T(\tau)$ with respect to $T$.
Since the latter quantity represents the energetic susceptibility of the system to the perturbation imposed by the bath,  we can interpret it as a generalized  Heat Capacity (HC) associated with the not-necessarily stationary state $\rho(\tau)$ of ${\cal S}$~\cite{Landau80,Falcioni11,Dicastro15}.
Irrespectively from the specific form of the probe/bath coupling, 
we can hence invoke   the QCRB to draw the following universal relation
\begin{eqnarray}
Q_T(\tau) \geq C_T^2(\tau)/\Delta^2 E_T(\tau)\;, \label{IMPO}
\end{eqnarray}   that links together the generalized HC of ${\cal S}$, its energy spread $\Delta^2E_T(\tau)$,  and the associated   QFI functional. 
The inequality~(\ref{IMPO})  can be shown to saturate at least in those cases where the  ETF  holds, i.e. when 
 $\tau$ is sufficiently long to ensure that, via thermalization, 
${\cal S}$ reaches the equilibrium state represented by the
thermal Gibbs state 
$\rho_{T}^{(eq)}=e^{-{\cal H}_S / k_BT}/\cal{Z}$,  with ${\cal Z}=\mbox{Tr} [e^{-{\cal{H}_S} / k_BT}]$ the partition function of the system. 
In this scenario in fact one has~\cite{ZANARDI1,ZANARDI2}
\begin{eqnarray} 
Q_{T}^{(eq)} = \frac{\Delta^2 E_T^{(eq)}}{k_B^2 T^4} \;, 
\qquad C_{T}^{(eq)} = \frac{\Delta^2 E_T^{(eq)}}{k_B T^2} \;,
\label{QuantumFisher111}
\end{eqnarray}
which indeed implies $Q_{T}^{(eq)} = [C_{T}^{(eq)}]^2/\Delta^2 E_T^{(eq)}$.
Accordingly one can conclude that, when the thermometer and the bath reaches thermal equilibrium, the standard
thermometry procedure which derive $T$ from the mean energy of ${\cal S}$, is optimal.  
We point out that Eq.~(\ref{QuantumFisher111}) also establishes a direct linear dependence between QFI and the associated capacity, i.e. 
\begin{eqnarray} 
Q_{T}^{(eq)} ={C_{T}^{(eq)}}/({k_B T^2})\;,
\label{QuantumFisher1110}
\end{eqnarray}
which, as we shall clarify in the following, is a peculiar property of  Gibbs states. 

\begin{figure}[t]
\includegraphics[width=1\columnwidth]{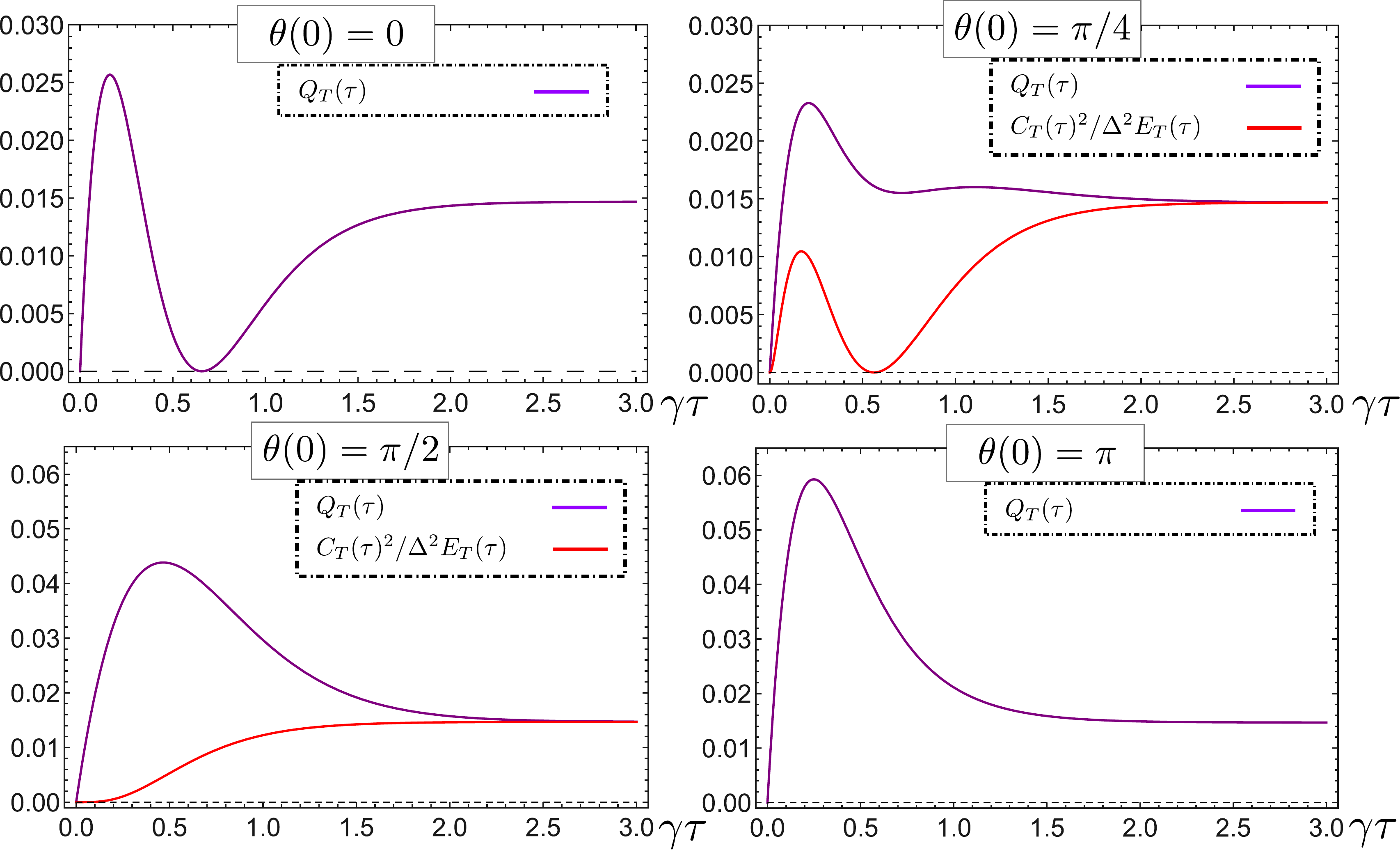}
\caption{(Color online) Plots of  the non-equilibrium QFI $Q_T(\tau)$ (purple curves)  and of the quantity $C_T^2(\tau)/\Delta^2 E_T(\tau)$ (red curves) 
appearing on the r.h.s. of Eq.~(\ref{IMPO}) and which determines the accuracy of the estimation procedure based on direct energy
measurement of the probe $\cal S$.  
In all plots we assume the input state $\rho(0)$ of ${\cal S}$ to be pure with zero azimuthal angle and with  polar angle given by $\theta(0)=0$ 
(excited state) for the first panel; $\theta(0)=\pi/4$ for the second; $\theta(0)=\pi/2$ for the third; and $\theta(0)=0$ (excited state)
for the last panel.
 The temperature $T$ is set equal to  $2$ in units of $\hbar \omega/k_B$ while the time  is measured
in units of $\gamma^{-1}$. 
Notice that when the system is initially prepared in a diagonal 
state, {\it i.e.} for $\theta(0)=0$ and $\theta(0)=\pi$, the bound (\ref{IMPO}) is saturated
and the two curves coincide.
}
\label{TheoreticalCurves}
\end{figure}
\paragraph{The Qubit model:--} Let us now focus on the
special case where the probe system ${\cal S}$ 
is a qubit with fixed Hamiltonian ${\cal H}=\hbar \omega \sigma_3 / 2$, and ${\cal B}$ is a Bosonic
thermal bath (hereafter $\sigma_3$ being the third Pauli operator). As in Refs.~\cite{Depasquale17,Campbell117,Brunelli11,Brunelli12,Correa15}
  we describe  the temporal evolution of ${\cal S}$ by assigning a Master Equation (ME)
which we write in the interaction picture representation as
 $\dot{\rho}(t) =
\sum_{j=\pm} \gamma_j {\cal D}_j[\rho(t)]$.
In this expression  ${\cal D}_-$ and ${\cal D}_+$ are Gorini-Kossakowski-Sudarshan-Lindblad (GKSL) generators having, respectively, the qubit ladder matrices
$\sigma_-= |0\rangle\langle 1|$ and $\sigma_+ = |1\rangle\langle 0|$ 
as corresponding  Lindblad operators  (hereafter $|0\rangle$ and $|1\rangle$ identify respectively the excited and the ground 
state of the single-qubit thermometer). 
The parameters  $\gamma_-=\gamma(N+1)$ and $\gamma_+=\gamma N$ instead 
set the temperature dependence of the system dynamics through the Planck number 
$N= 1/(e^{\hbar \omega/ k_B T} -1)\in[0,\infty[$  that counts the average 
number of resonant Bosonic excitations present in the bath, $\gamma$ being  a positive rate that fixes the
time scale of the problem. 
By direct integration of the ME one can easily verify that  the state of ${\cal S}$ at time $\tau$ can be expressed as
  $\rho(\tau)= \frac{1}{2} [ \openone + \vec{r}(\tau) \cdot 
\vec{\sigma}]$ with a Bloch vector 
$\vec{r}(\tau)$ having cartesian components equal to 
${r}_{1,2}(\tau) = {r}_{1,2}(0) e^{-\gamma(2 N+1) \tau/2}$ and ${r}_3(\tau) = {r}_3(0) e^{-\gamma(2 N+1) t} - 
(1- e^{-\gamma(2 N+1) \tau} )/(2N+1)$. This corresponds to an evolution induced by a Generalized Amplitude Damping (GAD) 
channels $\Phi_\tau$~\cite{Nielsen00} which, 
 irrespectively from the specific choice of  $\rho(0)$  will let the system to asymptotically 
 relax to a unique fixed point with  Bloch vector $\vec{r}^{(eq)}= (0,0,-1/(2N+1))$ which represents
 the system  thermal Gibbs state 
$\rho_{T}^{(eq)}$. In this long time limit, our model will behave as anticipated in the previous section, 
saturating the inequality (\ref{IMPO}), i.e.  allowing to recover the QCRB via ETF -- as well as fulfilling (\ref{QuantumFisher1110}).
 What about the finite time $\tau$ regime? 
 For the present model the heat capacity $C_T(\tau)$ and  the energy spread $\Delta^2 E_T(\tau)$ can be easily shown to be equal to  
 \begin{eqnarray} 
 C_T(\tau) = \tfrac{\hbar\omega}{2} \; \partial_T  r_3(\tau)\;, \quad 
 \Delta^2 E_T(\tau)= \left(\tfrac{\hbar\omega}{2}\right)^2
  [1 - r^2_3(\tau)]. 
 \end{eqnarray} 
Furthermore the   QFI can be computed as   
 $Q_{T}^{(\tau)}={\mbox{Tr}}[  L_T\; \partial_T \rho(\tau)]$ with $L_T$  being the (possibly time dependent) 
 Symmetric Logarithmic Derivative of the problem, i.e.  the self-adjoint operator which satisfies the identity $\partial_T \rho(\tau)=1/2 \; \lbrace L_T, \rho(\tau) \rbrace$, with $\{ \cdots, \cdots\}$ being the anti-commutator~\cite{Paris09}.
   Simple algebra allows us to express this as
\begin{equation}
Q_{T}{(\tau)} = \frac{[\partial_T \; {{r}(\tau)} ]^2}{1-{r^2(\tau)}} +{{r^2(\tau)} \; [\partial_T \theta({\tau})]^2}\;,
\label{QuantumFisher}
\end{equation}
where   ${r(\tau)}$ and  $\theta(\tau)$ are, respectively,  the length  and the polar angle of the Bloch vector $\vec{r}(\tau)$, the azimuthal angle being a constant of motion and playing no role in the derivation -- see Appendix for details.
The first term on the r.h.s. of Eq.~(\ref{QuantumFisher})  describes the rearrangement of the population of the probe during its interaction with the reservoir, while the other one accounts for quantum coherence contributions  which 
 nullifies in the asymptotic limit where $\gamma \tau \rightarrow \infty$ (the first term converging instead to $Q_{T}^{(eq)}$).
By direct substitution of these expressions into~(\ref{IMPO}) one can verify that for generic choices of $\tau$ and of the input state $\rho(0)$ the inequality will be strict -- see Fig.~\ref{TheoreticalCurves}. A notable exception however is obtained when the input state is diagonal into the 
energy basis of $H$, i.e. when $r_{1,2}(0)$ both nullify (or equivalently when, independently from the choice of $\rho(0)$,  the coherence terms of  $\rho(\tau)$ 
are removed by a decoherence process that acts on $\cal S$ before the measurement stage).
 In this special cases 
the system remains diagonal along the full trajectory and Eq.~(\ref{QuantumFisher}) reduces to 
$Q_{T}{(\tau)} = \frac{[\partial_T \; {{r}_3(\tau)} ]^2}{1-{r_3^2(\tau)}}$. Accordingly~(\ref{IMPO}) becomes an identity for all choices of
 the interaction time $\tau$, implying that the standard thermometry scheme which recovers $T$ from just energy measures is
 optimal. Notice that in this scenario, $\rho(\tau)$ has not reached the thermal equilibrium configuration so ETF arguments cannot be applied: this is made evident by the fact that even though (\ref{IMPO}) saturates, yet $Q_T(\tau)$ and $C_T(\tau)$ cannot be linearly connected as in (\ref{QuantumFisher1110}) unless one introduces an effective, yet fictitious,
 rescaling of the proportionality coefficient  appearing on the right-hand-side. 

The numerical plots of Fig.~\ref{TheoreticalCurves} show the relations between the l.h.s. and r.h.s terms of (\ref{IMPO}).
In agreement with the finding of Ref.~\cite{Correa15}
we notice that in general the QFI reaches higher values (corresponding to better estimation accuracies) for finite (possibly dependent on $T$) values of $\tau$. Furthermore after having fixed the parameter $\tau$ at its best,  
the absolute best performance is obtained when initializing the qubit into the ground state  (see last panel of the figure) -- we have confirmed this result by numerical optimization of (\ref{QuantumFisher}),
as shown in details in the Appendix. 
The first and last panel of Fig. \ref{TheoreticalCurves} explicitly show
the saturation of Eq. (\ref{IMPO}) for diagonal states at all times $\tau$, while for generic input this is only possible 
when $\tau \rightarrow \infty$ since the system
asymptotically thermalize.

\begin{figure}[h]
\includegraphics[width=0.9\columnwidth]{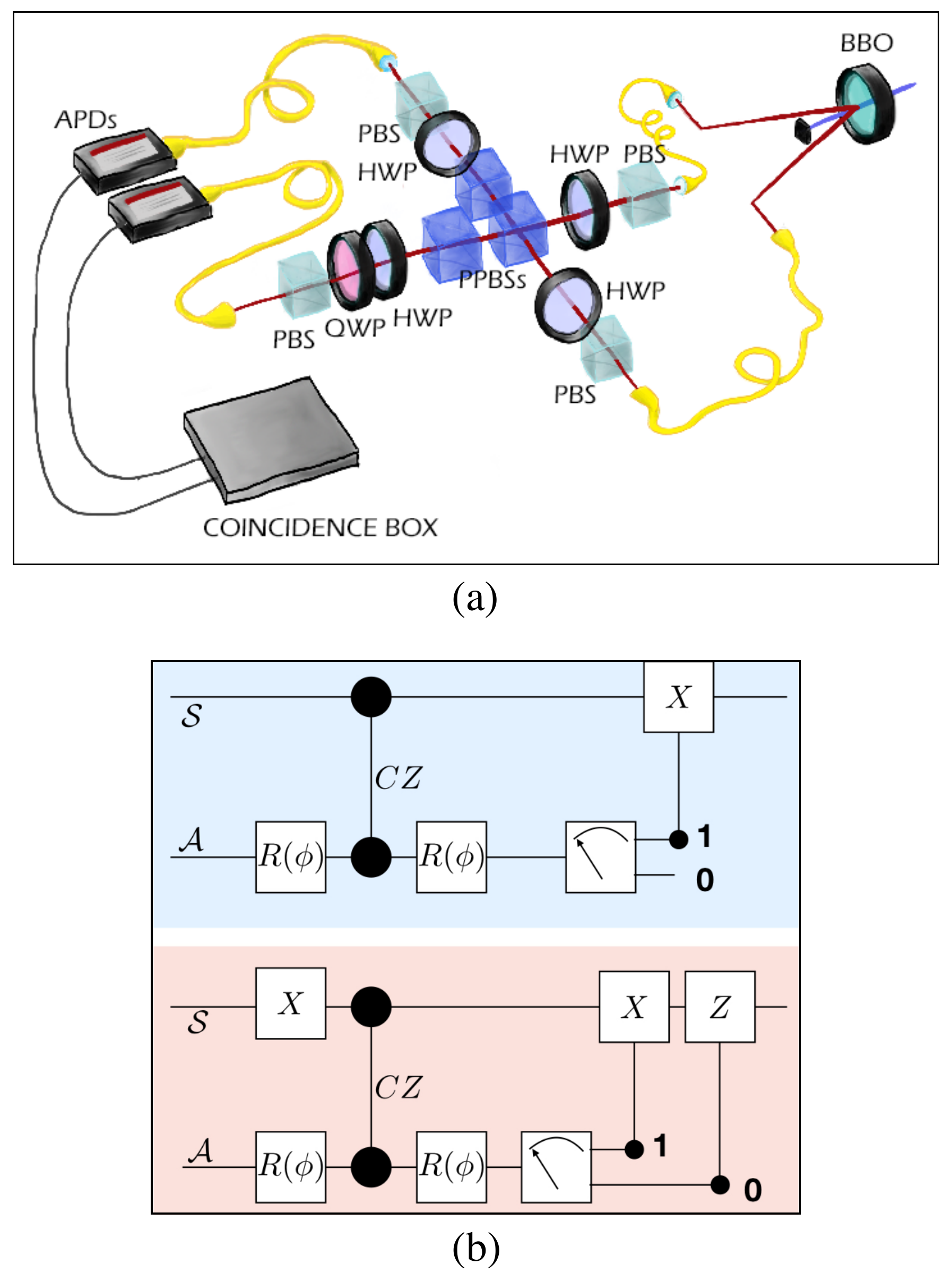}
\caption{Part (a): quantum simulation via quantum photonics. A photon pair is produced via a Spontaneous Parametric Down Conversion (SPDC) process through a Type-I 3 $mm$ BBO source. One photon is employed to simulate the single-qubit thermometer, while the other one is used as an ancilla 
to simulate the {\it system-bath} interaction. The computational basis is encoded in the vertical and the horizontal
polarizations of the single photons.
A Polarizing Beam Splitter (PBS) and a Half Wave Plate (HWP) on each arm
are used to prepare the state of the pair while the gate is composed by 
three Partially Polarizing Beam Splitters (PPBSs) \cite{Mancino17}. The final measure counts are 
collected using two Avalanche PhotoDiodes (APDs), and a Coincidence Box (FGPA). 
Part (b):
circuits for the simulation of the AD (top panel) and IAD (bottom panel) Channels \cite{Lu17}. The circuital elements are: X and Z, that 
implement the Pauli rotations $\sigma_x$ and $\sigma_z$; CZ, representing a controlled-$\sigma_z$ gate; $R(\phi)$ is a rotation by an angle $\phi$ around the $y$ axis. The measurements are performed in the computational basis.}
\label{ExperimentalSetup}
\end{figure}
\begin{figure*}[t]
\includegraphics[width=0.8\textwidth]{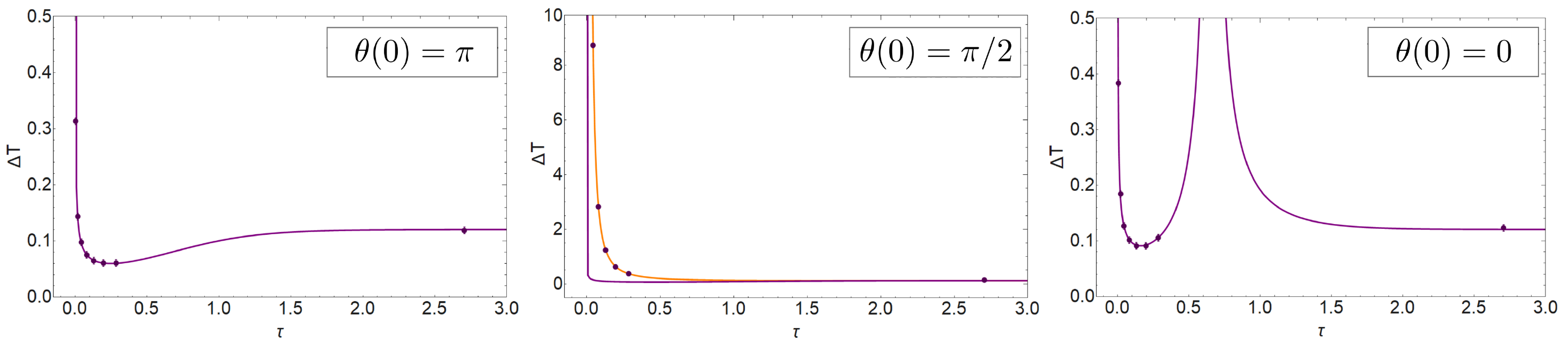}
\caption{Comparison between the experimental errors $\Delta T$, and metrological figures of merit related to the temperature parameter. In the three panels, the purple curves represent the theoretical QCRB, the orange curve represents the theoretical CRB, and the purple points represent the experimental uncertainties on the temperature. In the left panel, we confirm that the ground state allows to reach the greatest sensitivity of the single-qubit probe as it permits to reach the lowest value of $\Delta T$; in the right panel, we show the behaviour of the probe prepared in the excited state, and we observe a divergence in the QCRB due to the presence of a zero in the QFI --- see first panel of Fig.~\ref{TheoreticalCurves}; in the middle panel, we show the coherent strategy. Here, the experimental uncertainties on the temperature do not reach the QCRB but they are well captured by its classical counterpart.}
\label{ExperimentalResults}
\end{figure*}

\paragraph{Quantum Photonic Simulation:--} We have simulated the  evolution of the probing qubit ${\cal S}$  under the action of the thermal bath 
via a photonic implementation  of the associated GAD channel  $\Phi_t$~\cite{Wang13,Lu17,Aspuruguzik12, Cialdi2017}, in order to extract the experimental uncertainties on temperature estimation.
For this purpose we have exploited the Kraus representation of the map 
$\rho(\tau)=\Phi_{\tau} [\rho(0)]=\sum_{i=1}^4 K_{i}  \rho(0) {K_{i}}^\dagger$, where
 $K_i$'s are four Kraus operators:  the first two, i.e. 
$K_1=\sqrt{\tfrac{N+1}{2N+1}} (\vert 0 \rangle \langle 0 \vert + {e^{-\gamma(2N+1) \tau/2}} \vert 1 \rangle \langle 1 \vert)$, $K_2=
\sqrt{\tfrac{N+1}{2N+1}} \sqrt{1-e^{-\gamma(2N+1)  \tau}} \vert 0 \rangle \langle 1 \vert$, 
being responsible for decay from the excited to the ground state 
represent the action of an amplitude damping (AD) map, 
 the second two,
i.e. 
$K_3= \sqrt{\tfrac{1}{2N+1}}( e^{-\gamma(2N+1)  \tau/2} \vert 0 \rangle \langle 0 \vert + \vert 1 \rangle \langle 1 \vert)$, and $K_4=
\sqrt{\tfrac{1}{2N+1}}\sqrt{1-e^{-\gamma(2N+1)  \tau}} \vert 1 \rangle \langle 0 \vert$,
describing the 
absorption events, represent instead an inverse amplitude damping (IAD) map. 
The previous decomposition depicts the GAD as a weighted sum of two different processes, an AD and an IAD with 
weights respectively equal to 
$\sqrt{\tfrac{N+1}{2N+1}}$ and $\sqrt{\tfrac{1}{2N+1}}$.
This last property is crucial for implementing a quantum optical simulation of the process:
after reproducing the AD and the IAD channel through a succession of optical logic gates, is possible to reconstruct the full density matrix simply doing a proper weighted 
sum of the outputs of the two channels \cite{Mancino17}.
Specifically, an AD acting on a qubit $\mathcal{S}$ can be formally simulated by coupling the system with an ancilla $\mathcal{A}$ and doing the following operations : 
\begin{enumerate} 
\item a controlled-$\sigma_z$ gate, with $\mathcal{S}$ as the control, embedded between two rotations $R(\phi)$
acting on $\mathcal{A}$.
The rotations are performed around the $y$ axis and the angle $\phi$ 
has to be choosen in order to mimic the damping factor of the Kraus decomposition of the map, and in our case is such that $e^{-\gamma(2N+1) \tau} = \cos^2(2 \phi)$ 
\cite{Kiesel05, Mancino17, Mancavina18};
\item a projective measurement on the computational basis of $\mathcal{A}$, 
conditioning a $\sigma_x$ gate on $\mathcal{S}$ (see Fig. \ref{ExperimentalSetup} (b), top panel).
\end{enumerate}
The above mentioned procedure works also for the IAD, except for two additional $\sigma_x$ and $\sigma_z$ rotations in the preparation and post-processing of the state
(Fig. \ref{ExperimentalSetup} (b), bottom panel).
An experimental implementation is obtained by associating 
each logical gate with its corrispective element in the optical table, as explained in
Fig. \ref{ExperimentalSetup} (a).

The mean value of the energy and the temperature uncertainty 
are inferred performing a measure on the Hamiltonian eigenbasis of $\mathcal{S}$,
a purpose that in practice is realized through experimental counts of the populations \cite{Nota1}.
The expectation value of the energy is given by
$\langle E \rangle= (n_0-n_1)/2(n_0+n_1)$, where $n_i$ corresponds 
to the measured count rate of the state $i$. Its uncertainty is evaluated as 
$\Delta^2 E=n_0 n_1/(n_0+n_1)^3$; temperature uncertainties (at each estimation round) are then obtained 
as $\Delta^2 T = \Delta^2 E / (\partial_T E)^2$. The results are summarized in
Fig.~\ref{ExperimentalResults}, in which we compare the experimental uncertainties 
on the temperature with the related QCRB.

\paragraph{Conclusions:--} 
Wherever thermal equilibrium is reached, the ETF establishes a neat link between the temperature fluctuations $\Delta T$, and the thermal susceptibility of the system corresponding to the heat capacity. 
We have investigated whether  inspired relations can be recovered in non-equilibrium regimes. Studying the case of a single-qubit thermometer, we have explicitly shown that this is not possible whenever coherence is present in the initial state of the probe, as the QFI functional which gauges the optimal accuracy threshold contains additional contributions. However for diagonal
input states the optimality of standard measurement procedure is restored and allows to saturate the QCRB with conventional thermometry approaches based on
energy measurements. 
This peculiar effect is probably related with the small number of degree of freedom characterizing the thermometer we used. 
As a matter of fact, we suspect that as the dimensionality of the probing system increases, optimal thermometry could only be
achieved by more complex measurement procedures which, even in the absence of off-diagonal terms, include the study of the full 
statistic of the energy measures. 

{\it Acknowledgements.} 
ADP acknowledges financial support from the University of
Florence in the framework of the University Strategic Project
Program 2015 (project BRS00215).

\clearpage
\newpage
\widetext 

\appendix
\section{Derivation of equation (\ref{QuantumFisher})}
\label{Appa}
A convenient way to compute the QFI is to express  the symmetric logarithmic derivative $L_T$ operator of the problem in the Pauli basis, i.e. 
$L_{T} = l_0(\tau) \identity + \vec{l}(\tau) \cdot \vec{\sigma}$, with $l_0(\tau)$ and $\vec{l}(\tau)=(l_1(\tau),l_2(\tau),l_3(\tau))$ being real quantities
which can in principle depend upon the evolution time $\tau$. With this we can now write the identity $\{L_{T}(\tau), \rho(\tau)\} = 2 \partial_T \rho(\tau)$ as 
\begin{eqnarray} && l_j (\tau) + r_j(\tau)  l_0 (\tau)  =  \partial_T r_j(\tau) \;, \qquad j=1,2,3\;, \label{2app}\\
&&l_0 (\tau) + \sum_{j=1}^3 r_j(\tau)  l_j(\tau)=0\;, \label{3app}
\end{eqnarray} 
with $r_j(t)$ being the cartesian components of the  the Bloch vector $\vec{r}(t)$ of $\rho(\tau)$. 
By substitution of Eq.~(\ref{2app}) into Eq.~(\ref{3app}) we obtain
\ba \begin{gathered}
l_0 (\tau) = -\frac{1}{2} \frac{\partial_T {r}^2(\tau) }{1-{r}^2(\tau)}\;, 
\end{gathered}  \ea
with $r(\tau) = \sqrt{ \sum_{j=1}^3 r^2_j(\tau)}$ being the length of $\vec{r}(\tau)$. Replacing this  into (\ref{2app}) we can then write  
\ba \label{4app} \begin{gathered} l_j(\tau)   =  \partial_T r_j(\tau) + \frac{r_j(\tau) }{2} \frac{\partial_T {r}^2(\tau) }{1-{r}^2(\tau)} \;,
\end{gathered} \ea
and hence 
\ba Q_T = \mbox{Tr}[\partial_T \rho L_{T}] = \sum_{j=1}^3 l_j(\tau) \partial_T r_j(\tau) 
= \sum_{j=1}^3 \big[  \partial_T r_j(\tau)\big]^2 + 
\frac{1}{4} \frac{\big[\partial_T r^2(\tau)  \big]^2}{1 - r^2(\tau)} = \sum_{j=1}^3 \big[  \partial_T r_j(\tau)\big]^2 + 
r^2(\tau)  \frac{\big[\partial_T r(\tau)  \big]^2}{1 - r^2(\tau)}\;. \label{FFDS} 
 \ea
Expressing then the Bloch vector in polar coordinates  $\vec{r}= r (\cos\phi\sin\theta, \sin\phi \sin\phi, \cos\theta)$  we notice that 
 the system ME
admits the azimuthal angle as constant of motion, i.e. 
$\phi(\tau) = \arctan[r_2(\tau)/r_1(\tau)] = \arctan[r_2(0)/r_1(0)] =\phi(0)$, which, by construction cannot depend upon $T$. Exploiting this fact it turns out that (\ref{FFDS}) only depends upon 
the partial derivative in $T$ of the modulus $r(\tau)$ and of the polar angle $\theta(\tau)$  as shown in (\ref{QuantumFisher}).

\section{QFI for a qubit in a bosonic channel}
\label{Appb}

\begin{figure}[t]
\includegraphics[width=0.8\textwidth]{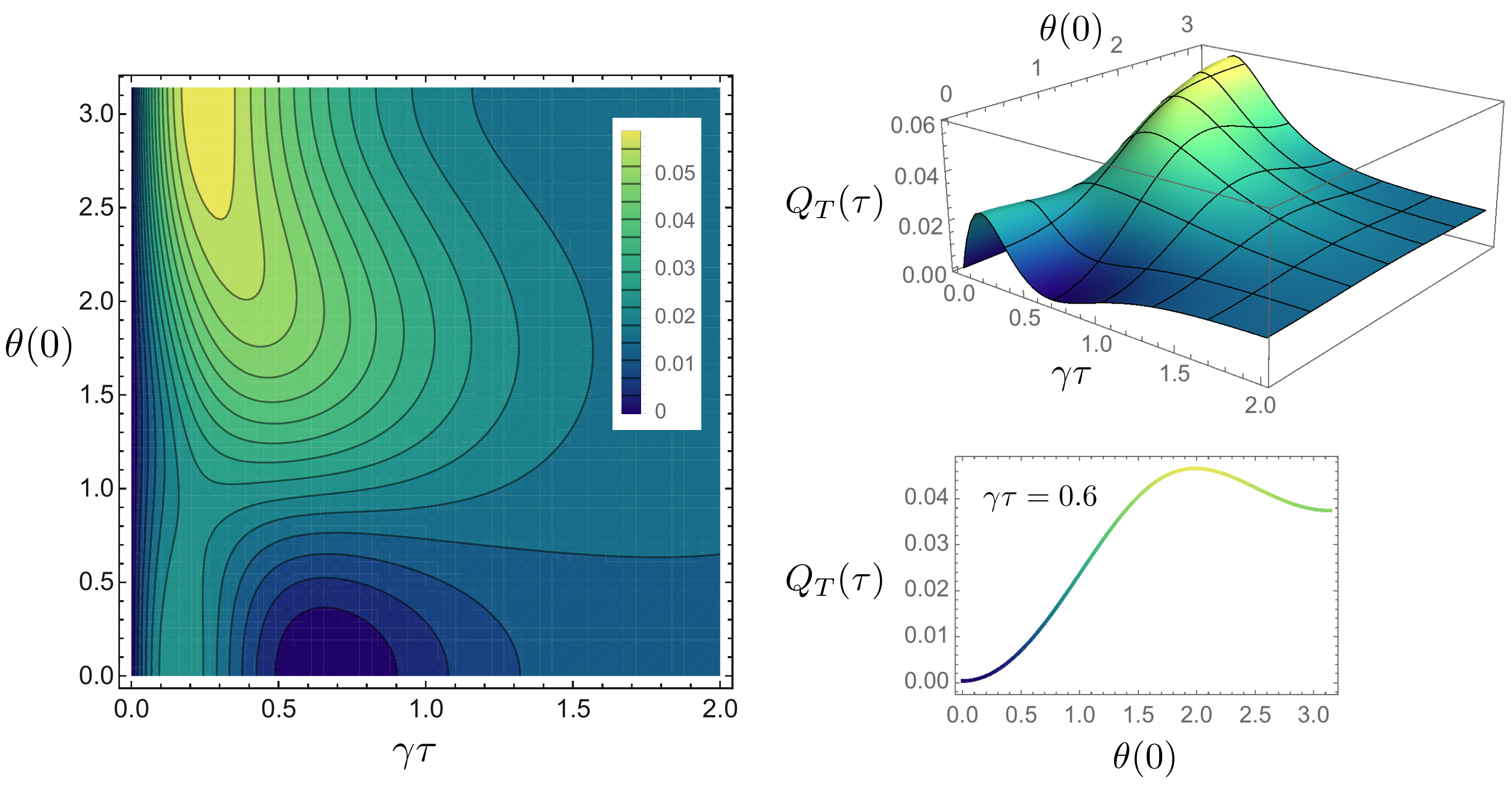}
\caption{Contour-plot (left panel) and 3D plot (upper right panel) of the QFI~(\ref{QuantumFisher})  of the thermalizing probe as a function of the dimensionless time $\gamma \tau$ and 
the polar angle $\theta(0)$ of the initial state. 
Lower right panel: 
the QFI at fixed $\gamma \tau = 0.6$ for different initial preparations $\theta(0)$, highlighting that the 
fundamental state is not the optimal choice in this case. In all the plots the temperature is set to 
$2$ in units of $\hbar \omega/k_B$.}
\label{Total}
\end{figure}

The value of the QFI for a two level system evolving through the 
GAD considered in the main text is represented in Fig.~\ref{Total} 
for different times and initial preparations.
The plot shows that initialising the probe in the fundamental state
is the optimal choice for temperature estimation, and in agreement with 
\cite{Correa15} the best performance is attained waiting a finite
amount of time.
This particular behaviour can be explained  observing  that the 
decay rate of the populations is explicitly dependent
by the average number of resonant bosonic excitations, and consequently contains some information
about the temperature, that is eventually lost if the system 
achieves complete thermalization.
In this last scenario the QFI becomes independent on the initial conditions, as 
it is clearly shown in the upper right corner of Fig. (\ref{Total})
and its value asymptotically satisfies Eq. (\ref{QuantumFisher1110}) that 
holds for thermalized probes.
Notice that the additional dependence on $T$ provided by the decay
rate is not always an advantage for temperature estimation, as it is
evident from the low-$\theta(0)$ region of the contour-plot and from the last panel 
of Fig. (\ref{TheoreticalCurves}), that displays a null QFI
for a probe initialized in the excited 
state for a properly chosen time of measurement.
Finally we remark that the fundamental state is no longer optimal if we fix different
values of the intermediate time $\tau$, as pointed out for instance in the
lower right corner of Fig. (\ref{Total}) in which the theoretical curve
for $\gamma \tau=0.6$ is represented.

\end{document}